# 面向 6G 的无线通信信道特性分析与建模

王承祥[1,2]，黄杰[1,2]，王海明[2,3]，高西奇[1,2]，尤肖虎[1,2]，郝阳[4]

（1. 东南大学信息科学与工程学院移动通信国家重点实验室，江苏 南京 210096；2. 紫金山实验室，江苏 南京 211111；
3. 东南大学信息科学与工程学院毫米波国家重点实验室，江苏 南京 210096；
4. 伦敦玛丽女王大学电子工程与计算机科学学院，英国 伦敦 E1 4NS）

**摘　要：** 针对 6G 全覆盖、全频谱、全应用的发展愿景，对面向 6G 的全频谱全场景无线信道测量、信道特性与信道模型方面的进展进行了全面概述，侧重于毫米波、太赫兹、光波段、卫星、无人机、海洋、水声、高铁、车对车、大规模/超大规模天线、轨道角动量以及工业物联网等通信信道，并展示了 6G 信道的相关测量与建模结果。最后，指出了 6G 无线信道测量与建模研究的未来挑战。

**关键词：** 6G 无线通信网络；信道测量；信道特性；信道建模；信道模型性能评估

# 6G oriented wireless communication channel characteristics analysis and modeling

WANG Chengxiang[1,2], HUANG Jie[1,2], WANG Haiming[2,3], GAO Xiqi[1,2], YOU Xiaohu[1,2], HAO Yang[4]

1. National Mobile Communications Research Laboratory, School of Information Science and Engineering, Southeast University, Nanjing 210096, China
2. Purple Mountain Laboratories, Nanjing 211111, China
3. State Key Laboratory of Millimeter Waves, School of Information Science and Engineering, Southeast University, Nanjing 210096, China
4. School of Electronic Engineering and Computer Science, Queen Mary University of London, London E1 4NS, U.K.

**Abstract:** Based on the vision on the 6G wireless communication network, i.e., global coverage, all spectrums and all applications, we comprehensively survey 6G related wireless channel measurements, channel characteristics, and channel models for all frequency bands and all scenarios. Millimeter wave (mmWave), terahertz (THz), optical band, satellite, unmanned aerial vehicle (UAV), maritime, underwater acoustic, high-speed train (HST), vehicle-to-vehicle (V2V), massive/ultra-massive multiple-input multiple-output (MIMO), orbital angular momentum (OAM), and industry Internet of things (IoT) communication channels were particularly investigated. The related 6G channel measurement and modeling results were also given. Finally, future research challenges on 6G channel measurements and modeling were pointed out.

**Key words:** 6G wireless communication network, channel measurement, channel characteristic, channel modeling, performance evaluation of channel model

## 1 引言

从 2020 年起,5G 无线通信网络已在全球范围内进行部署。5G 的应用场景包括增强型移动宽带(eMBB, enhanced mobile broadband)、大规模机器类通信(mMTC, massive machine type communication)与高可靠低时延通信(uRLLC, ultra-reliable and low latency communication),应用需求是实现移动互联与万物互联。但是,5G 并不能满足未来无线通信网络的所有需求,面向 6G 无线通信网络的相关研究已经迅速展开[1-8]。从应用需求的角度看,潜在的 6G 无线通信网络将在 5G 的基础上继续深化移动互联,不断扩展万物互联的边界和范围,最终实现万物智联。

应用需求驱动技术需求。从峰值传输速率的角度看,6G 将在 5G 的几十 Gbit/s 的基础上提高数十倍,达到 Tbit/s 级的超高传输速率(巨流量);从连接数密度的角度看,6G 将在 5G 的 100 万/km² 的基础上提高 10~100 倍,达到 1 000 万/km²~10 000 万/km²(巨连接);从覆盖率的角度看,6G 将在 5G 陆地局部覆盖的基础上实现全球深度覆盖(广覆盖);从智能化的角度看,6G 将由 5G 时代的初步智能最终实现高度智能(高智能)。5G 与 6G 应用需求及技术需求对比如表 1 所示。

表 1 5G 与 6G 应用需求及技术需求对比

| 无线通信网络 | | 5G | 6G |
|---|---|---|---|
| 应用需求 | | 移动互联 万物互联 | 深化移动互联 扩展万物互联+万物智联 |
| 技术需求 | 峰值传输速率 | 几十 Gbit/s | Tbit/s(巨流量) |
| | 连接数密度 | 100 万/km² | 1 000 万/km²(巨连接) |
| | 覆盖率 | 陆地局部覆盖 | 全球深度覆盖(广覆盖) |
| | 智能化 | 初步智能 | 高度智能(高智能) |

6G 的应用场景极为丰富,将进一步强化 5G 的 eMBB、mMTC、uRLLC 三大应用场景,并扩展到空天地海一体化网络、人工智能使能网络等应用场景。受新的应用需求与技术需求的驱动,6G 需要引入新的性能指标,如更高的频谱效率/能效/成本效率、更高的传输速率、更低的时延、更大的连接数密度、覆盖率、智能化程度、安全性等。5G 的峰值速率为 20 Gbit/s,而 6G 由于太赫兹及光波段频段的应用,其峰值速率将达到 1~10 Tbit/s。由于高频段的应用,用户体验速率将达到 Gbit/s 级,流量密度将超过 1 Gbit/(s·m²)。由于人工智能的应用,与 5G 相比,6G 的频谱效率将提高 3~5 倍,能效将提高 10 倍左右。由于超密集异构网络、空天地海多种通信场景、多种通信频段(6 GHz 以下到光波段)的应用,连接数密度将提高 10~100 倍。由于超高速铁路、无人机、卫星的应用,移动速度将支持 1 000 km/h 以上。网络的端到端时延将小于 1 ms。并且,为了对 6G 网络进行更全面的评估,需要引入其他重要的性能指标,如成本效率、覆盖率、智能化程度和安全容量等。

为了满足以上应用需求与性能指标,6G 无线通信网络将采取新的范式并依赖于新的使能技术。新的范式可以总结为全覆盖、全频谱、全应用、强安全四大趋势。为了提供全球覆盖,6G 无线通信网络将从陆地移动通信扩展到空天地海一体化通信网络,包括卫星通信网络、无人机通信网络、陆地超密集网络、地下通信网络、海洋通信网络以及水声通信网络。为了满足超高传输速率(巨流量)和超高连接密度(巨连接)的应用需求,包括 6 GHz 以下、毫米波、太赫兹、光波段在内的全部频谱将被充分挖掘。为了满足全应用的需求,人工智能和大数据将与无线通信网络全面融合以更好地管理通信网络[9]。并且,人工智能可以更好地动态编排网络、缓存与计算资源以提高系统性能。强安全体现在设计网络的同时,将安全因素考虑在内,也称为网络内生安全,包括物理层安全与网络层安全两个方面。

无线信道是系统设计、网络优化、性能评估的基础。为了实现 6G 无线通信网络,亟需对 6G 无线信道展开深入研究。本文对 6G 全频谱全场景无线信道测量、信道特性与信道模型方面的进展进行了全面概述,侧重于毫米波、太赫兹、光波段、卫星、无人机、海洋、水声、高铁、车对车、大规模/超大规模天线、轨道角动量以及工业物联网等通信信道,并展示了 6G 信道的相关测量与建模结果。同时,指出了 6G 无线信道研究方面的未来挑战,包括 6G 无线信道测量,基于智能反射面的 6G 技术的信道测量与模型,人工智能使能的信道测量与模型,6G 通用标准化信道模型,6G 信道模型参数、信道统计特性、通信系统性能之间的复杂映射关系,6G 信道模型的性能评估。

## 2 面向 6G 的无线信道测量与信道特性

面向 6G 的无线通信网络将是空天地海一体化

网络，包含多频段与多场景下不同类型的信道，主要有毫米波、太赫兹、光波段、卫星、无人机（空—空、空—地）、海洋（无人机—船、船—船、船—岸基）、水声、高铁、车对车、大规模/超大规模天线、轨道角动量以及工业物联网等通信信道。针对不同类型的信道，用于开展信道测量实验的信道探测器在测量原理、通道数、频段、带宽、发射功率、动态范围、移动性等方面具有较大差异。因为多种关键技术将在 6G 无线通信网络中综合运用，所以不同类型的信道既具有独特的信道特性，又具有一定的关联。对不同类型的面向 6G 的无线信道测量与信道特性进行了全面总结，如表 2 所示。接下来对不同类型的 6G 信道展开论述。

## 2.1　毫米波/太赫兹信道

毫米波是指 30～300 GHz 频段的电磁波，由于 10～30 GHz 频段的电磁波具有与毫米波相似的电波传播特征，通常也将其考虑在内，而太赫兹是指 0.1～10 THz 频段的电磁波。因此，100～300 GHz 频段的电磁波兼具毫米波与太赫兹的一些信道特性，如大带宽、高定向性、大路径损耗、阻挡效应、大气衰减、漫散射等。毫米波用于几百米范围内 Gbit/s 量级的数据传输，而太赫兹可以支持几十米范围内 Tbit/s 量级的数据传输。相比于毫米波，太赫兹表现出更严重的路径损耗、大气衰减与漫散射现象。

毫米波信道已在一些典型频段得到充分研究，如 26 GHz/28 GHz、32 GHz、38 GHz/39 GHz、60 GHz、73 GHz 等频段，文献[10]对毫米波信道测量的最新进展进行了总结。尽管如此，多数现有的毫米波信道测量结果具有一定的局限性，主要体现在信号带宽（500 MHz 典型值）、距离（通常 300 m 以内）、场景（室内及室外静止环境）、天线配置（单天线）等方面。同时，由于不同的课题组使用不同的信道探测器、测量设置、测量场景和数据后处理方法，很难对不同频段的毫米波信道特性进行公平比较[11]。室外场景、多天线、高动态（如车对车）条件下的毫米波信道测量实验仍有待开展。28 GHz 毫米波车对车通信信道的归一化功率时延分布如图 1 所示。不同快拍的直射径及反射径功率在时域呈现连续变化，表明了毫米波车对车通信信道的空间连续性与时域非平稳性。对于 100 GHz 以上频段，文献[12]在 140 GHz 频段开展了初步的信道测量，对路径损耗、穿透损耗、散射特性等进行了研究。

对于太赫兹信道而言，多数的信道探测器基于矢量网络分析仪，外加不同太赫兹频段的上下变频器模块。常用的信道探测器为基于 220～325 GHz 频段的矢量网络分析仪及 220～330 GHz/260～400 GHz 频段的太赫兹上下变频模块。由于太赫兹频段的设备更昂贵且难以设计，发射功率与系统动态范围受

表 2　面向 6G 的无线信道测量与信道特性总结

| 面向 6G 的无线信道 | 频段 | 场景 | 信道特性 |
| --- | --- | --- | --- |
| 毫米波/太赫兹信道 | 26 GHz/28 GHz、32 GHz、38 GHz/39 GHz、60 GHz、73 GHz（毫米波），300 GHz 左右（太赫兹） | 室内、室外 | 大信道带宽，高定向性，大路径损耗，阻挡效应，大气衰减，漫散射 |
| 光波段信道 | 380～780 nm | 室内、室外、地下、水下 | 不同材料的复杂散射特性，收发端非线性光电特性，背景噪声 |
| 卫星通信信道 | Ku、K、Ka、V 波段 | 静止轨道、低/中/高轨道 | 雨/雪/云/雾衰减，极大的多普勒频移信道与多普勒扩展，大覆盖范围，长通信距离 |
| 无人机通信信道 | 2 GHz、2.4 GHz、5.8 GHz | 市区、郊区、农村、开阔场景（空—空与空—地） | 三维任意轨迹（大俯仰角），高移动性，空时非平稳性，机架阴影衰落 |
| 海洋通信信道 | 2.4 GHz、5.8 GHz | 无人机—船、船—船、船—岸基 | 散射体稀疏性，海浪运动影响，海洋表面波导效应，时变非平稳，长通信距离，气候影响 |
| 水声信道 | 2～32 kHz | 水下环境 | 大传输损耗，多径传播，时变非平稳，多普勒频移 |
| 高铁/车对车通信信道 | 6 GHz 以下，毫米波 | 开阔场景、山地、高架桥、隧道、路堑、车站、车厢内部（高铁）、高速公路、市区街道、开阔场景、校园、停车场（车对车） | 大多普勒频移与多普勒扩展，非平稳性，列车/车辆影响，速度与轨迹变化 |
| 大规模/超大规模天线信道 | 6 GHz 以下，毫米波，太赫兹 | 室内、室外 | 空间非平稳，信道硬化，球面波 |
| 轨道角动量信道 | 毫米波 | 直射、非直射（反射） | 复用增益，波束发散与失准，反射场景性能下降 |
| 工业物联网信道 | 6 GHz 以下 | 工业物联网场景 | 变化的路径损耗，随机波动，非直射传播，散射体丰富，多移动性 |

限,现有的太赫兹信道测量范围大多在桌面距离的量级或体域网[13]。多数太赫兹信道测量集中在 300 GHz 频段[14-16],300 GHz 以上频段的信道特性仍不明确,需要在未来开展大量的信道测量予以探究。

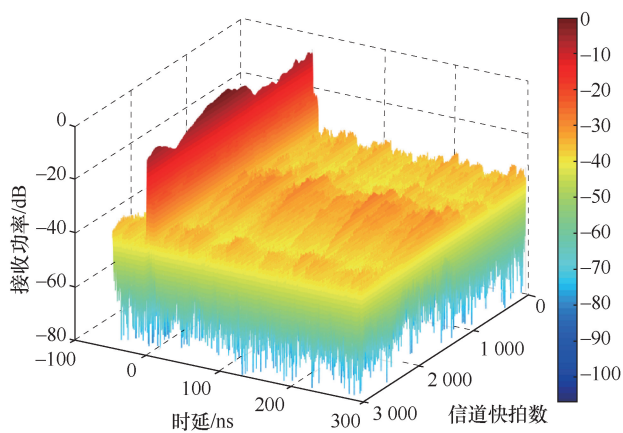

图 1 28 GHz 毫米波车对车通信信道的归一化功率时延分布

### 2.2 光波段信道

光波段是指载频为红外线、可见光、紫外线的电磁波谱,对应的波长分别为 780~$10^6$ nm、380~780 nm、10~380 nm[17]。光波段频段可用于室内、室外、地下、水下等场景的无线通信。

光波段信道展现出一些独特的信道特性,如不同材料的复杂散射特性、收发端非线性光电特性、背景噪声等。光波段信道场景可以分为直接视距、非直接视距、非直接非视距、追踪等。与传统无线电波信道相比,光波段信道没有多径衰落、多普勒效应以及频谱管制。测量的光波段信道特性主要包括信道冲激响应/信道传输函数、路径损耗、阴影衰落、均方根时延扩展等。文献[17]对光波段信道测量进行了详细总结。

### 2.3 卫星通信信道

卫星通信尽管已在导航、通信与广播领域使用多年,但仍然是现代无线通信的研究热点,并因可靠的服务质量与较低的成本被用于提供全球覆盖[18]。通常,卫星通信轨道可以分为静止轨道与非静止轨道。静止轨道卫星使用位于赤道上方 35 786 km 的对地同步卫星进行通信;根据卫星到地球的距离,非静止轨道可以进一步分为低轨道、中轨道、高轨道。卫星通信常用的频段包括 Ku(12~18 GHz)、K(18~26.5 GHz)、V(40~75 GHz)频段。

卫星通信信道受动态天气条件(如雨、云、雪、雾等)的影响极大,降雨是引起卫星信号衰减的主要因素,尤其是对 10 GHz 以上频段的电磁波来说。此外,卫星通信信道表现出极大的多普勒频移与多普勒扩展、频率相关性、大覆盖范围、长通信距离等特点。由于通信距离极长,卫星通信信道可以认为是视距传输,多径效应可以忽略。同时,需要使用大发射功率与高增益天线以对抗由于长距离和高频段引起的大路径损耗。

### 2.4 无人机通信信道

近年来,无人机在民用与军用领域发展迅速。无人机通信信道展现出一些新的信道特性,如三维部署、高移动性、空时非平稳性、机架阴影影响等[19-21]。

通常,无人机通信信道可以分为空—空与空—地信道。小/中型载人飞机与无人机常用于信道测量实验中。使用载人飞机进行信道测量的成本较大,而使用无人机可以极大地降低测量成本。现有文献同时开展了窄带与宽带信道测量,多数测量活动集中在 2 GHz、2.4 GHz、5.8 GHz 频段。测量场景包括市区、郊区、农村与开阔场景。测量的信道特性包括路径损耗、阴影衰落、均方根时延扩展、莱斯因子、信号幅度的概率密度分布/累计概率分布等。

### 2.5 海洋通信信道

由于蓝色经济的发展,海洋通信受到人们极大地关注。作为空天地海一体化网络的一部分,海洋通信信道主要包括空—海与近海表面信道[22]。对于空—海信道,无人机或中继站被用作空中基站,与海洋表面的船只进行通信,这类信道也称为无人机—船信道。对于近海表面信道,船只可与其他船只(船—船)或靠近海岸的基站(船—岸基)进行通信。海洋传播的独特性带来新的信道特性,如散射体稀疏性、海浪运动影响、海洋表面波导效应、时变非平稳、长通信距离、气候影响等。

已有海洋通信信道测量在 2.4 GHz 和 5.8 GHz 等频段开展,最远距离达 10 km。测量结果表明,船—船、船—岸基、波动、漂移情况下的路径损耗不同。对于无人机—船信道,主要存在直射径与反射径,均方根时延扩展较小。如果考虑直射径与一条反射径之外的第三条径,海浪及管道层将对均方根时延扩展产生较大的影响。随着距离的增加,莱斯因子趋于增加。

### 2.6 水声信道

水声信道面临很多挑战。由于海洋的背景噪声大,水声通信的可用频段很低,传输损耗较大[23]。同一水平面内的水下信道受折射、反射、散射等多

径传播的影响。此外，水声信道在时域与频域弥散，产生信道的时变特性与多普勒效应。信道测量通常在几 kHz 频段开展，分布在 2～32 kHz[24]。

### 2.7 高铁/车对车通信信道

高铁作为一种快捷、环保、灵活的交通方式，已在世界范围内广泛部署。之前的高铁通信系统主要为 GSM-R 和 LTE-R。最近，5G 网络被应用于高铁通信以提高信号服务质量。未来的超高速铁路速度有望突破 500 km/h，由此带来了频繁快速的小区切换和大多普勒扩展。毫米波/太赫兹与大规模/超大规模天线是用于提高高铁通信系统传输速率的潜在关键技术。一些初步的信道测量工作已在多种高铁通信场景中开展，包括开阔场景、山地、高架桥、隧道、路堑、车站、车厢内部等[25]。

车联网是 5G/6G 网络在高可靠低时延通信场景下的垂直行业典型应用，其信道类型包括车对车、车对基础设施、车对行人，可统称为 V2X。6 GHz 以下的车对车通信信道已有广泛研究，而毫米波频段的车对车通信信道测量结果较少。文献[26]对现有的毫米波车对车通信信道测量结果进行了总结。总的来说，在 28 GHz、38 GHz、60 GHz、73 GHz、77 GHz 频段开展了车对车通信信道测量实验。所有测量都采用两端单天线配置，其中一端可以固定或移动，天线放置于车内或车外，测量场景包括高速公路、市区街道、开阔场景、校园、停车场等。除此之外，配置多天线甚至大规模/超大规模天线的高移动性毫米波车对车通信信道测量实验亟需开展，如何使用高效率、低成本的方式对其进行测量仍有待研究。

### 2.8 大规模/超大规模天线信道

大规模/超大规模天线是配置有多达数百根天线的增强型多天线技术，它的出现是为了满足未来无线通信网络对于巨流量、巨连接的需求，能够大幅度提高系统容量和可靠性[27]。

大规模/超大规模天线信道特性主要包括空间非平稳、信道硬化与球面波特性。某些特定的簇不能被整个阵列天线都观察到（簇在阵列轴上是非平稳的），在阵列轴上将观察到簇的消失和出现。此外，在整个阵列上也将观察到功率的不平衡以及莱斯因子的变化，不同的天线单元之间的功率、时延参数也将发生漂移。信道硬化是指由于窄波束和大规模/超大规模天线阵列的应用，导致信道波动下降，用户之间的正交性越来越强。由于大规模/超大规模天线的应用，收发天线距离可能在瑞利距离以内，从而表现出球面波特性。文献[28-34]在 6 GHz 以下频段多种室内及室外场景开展了信道测量实验。

### 2.9 轨道角动量信道

轨道角动量表示电子绕传播轴旋转，能量流围绕光轴旋转产生，使得电磁波的相位波前呈涡旋状，因此，携带有轨道角动量的电磁波也被称为涡旋电磁波。轨道角动量已在多个研究领域引起广泛关注，尤其是无线通信领域。理论上看，单个波束内正交的轨道角动量模式的数目是无穷的，每个模式都是完备正交基中的一个基底，可以用于传输不同的信号以获得复用增益。根据电动理论，电磁波在传播的同时携带线性动量与角动量，其中，角动量包括自旋角动量与轨道角动量。

根据相关文献研究，轨道角动量信道具有一些新的信道特性。首先，轨道角动量复用与多天线复用具有一定的关联。文献[35]指出基于均匀圆阵的轨道角动量系统是多天线系统的子集，不会增加信道容量。文献[36]指出传统多天线匙孔信道性能较差，而基于均匀圆阵的轨道角动量系统在匙孔信道下则具有较好的性能。其次，轨道角动量波束的发散与失准会严重缩短传输距离，大特征值对应的传播模式具有更大的衰减。此外，信号的反射会破坏轨道角动量信道中涡旋电磁波的正交性，因而在非直射场景中较难检测到轨道角动量的不同模式。已有一些文献在毫米波频段开展了直射及非直射场景的轨道角动量信道测量实验[37-40]。

### 2.10 工业物联网信道

在工业物联网场景中存在大量的机器人、传感器、机械设备，这些设备之间需要以稳定、高效的方式进行海量连接。Sigfox、LoRa、NB-IoT 是 3 种常用的物联网标准，它们在功率消耗、频段、通信距离、传输速率、连接设备数等性能指标上有所不同。工业物联网信道展现出许多新的特性，如变化的路径损耗、随机抖动、非直射传播、丰富的散射体以及多移动性[41]。

目前，工业物联网信道测量的结果较少，这些信道测量主要集中在 6 GHz 以下频段[42]，与现有物联网标准使用的频段相同。然而，对未来物联网的巨流量与巨连接而言，毫米波频段的工业物联网信道测量极具前景[43]。

## 3 面向 6G 的无线信道模型

无线信道包括大尺度特性（路径损耗与阴影衰

落）和小尺度特性（多径）。通常，信道模型可以分为确定性模型与随机性模型两类。确定性信道模型包括基于测量的模型与射线追踪模型。基于地图的模型和点云模型是简化的射线追踪模型。随机性信道模型可以分为基于几何的信道模型与基于相关的信道模型。几何随机信道模型包括纯几何信道模型与半几何信道模型两类。纯几何信道模型又可分为规则形状与非规则形状两类，其中，非规则形状几何随机模型被大多数标准化信道模型所采用。文献[44]对 5G 信道模型进行了详细介绍。

由于不同的面向 6G 的无线信道具有独特的信道特性，研究人员采用多种建模方法，针对不同类型的无线信道提出了多种能够准确描述其信道特性的大尺度与小尺度信道模型。面向 6G 的无线信道模型总结如表 3 所示。

表 3　面向 6G 的无线信道模型总结

| 面向 6G 的无线信道 | 信道模型 |
|---|---|
| 毫米波/太赫兹信道 | 确定性：射线追踪模型、基于地图的模型、点云模型、准确定性模型 |
| | 随机性：SV 模型、传播图模型、几何随机信道模型 |
| 光波段信道 | 确定性：递归模型、迭代模型、DUSTIN 算法、天花板反射模型、几何确定信道模型 |
| | 随机性：几何随机信道模型、非几何随机信道模型 |
| 卫星通信信道 | 雨/云/雪/雾衰减模型、马尔可夫模型、几何随机信道模型 |
| 无人机通信信道 | 确定性：射线追踪模型、分析性模型 |
| | 随机性：规则形状几何随机信道模型、非规则形状几何随机信道模型、非几何随机信道模型、马尔可夫模型 |
| 海洋通信信道 | 确定性：射线追踪模型、两径模型、三径模型 |
| | 随机性：几何随机信道模型、两径衍射功率模型 |
| 水声信道 | 确定性：射线追踪模型 |
| | 随机性：瑞利/莱斯/对数正态分布模型 |
| 高铁/车对车通信信道 | 确定性：射线追踪模型 |
| | 随机性：几何随机信道模型、QuaDRiGa 模型、动态模型、马尔可夫模型、传播图模型 |
| 大规模/超大规模天线信道 | 几何随机信道模型 |
| 轨道角动量信道 | 暂无 |
| 工业物联网信道 | 路径损耗模型 |

## 3.1 毫米波/太赫兹信道模型

文献[10]对毫米波信道模型进行了全面总结。确定性信道模型包括射线追踪、基于地图的模型和点云模型。射线追踪模型应用于 IEEE 802.11ad 标准化模型中，而基于地图的模型由 METIS 项目组提出并应用于毫米波信道建模中。准确定性模型应用于 MiWEBA 项目组及 IEEE 802.11ay 标准化模型中。随机性信道模型包括 SV 模型、传播图模型、几何随机信道模型。几何随机信道模型应用于 NYU WIRELESS、3GPP 38.901、METIS、mmMAGIC 等模型中。射线追踪模型与几何随机信道模型也广泛应用于太赫兹信道建模中。同时，人体/植被衰减、雨/云/雪/雾衰减在毫米波及太赫兹频段也需要考虑。

太赫兹通信频率高，传输带宽大，会引发信道的频域非平稳性；由于路径损耗较大，收发两端通常需要采用大规模/超大规模天线阵列，会引发信道的空域非平稳性；另外，收发端设备的移动性会引发信道的时域非平稳性。针对太赫兹信道的信道特性，提出一种三维空—时—频非平稳室内信道模型。该模型使用几何随机建模方法，信道参数在空域、时域、频域演进以体现其非平稳特性。给出了不同时间（$t = 0, 10$ s）、不同频率（$f = 300$ GHz, 350 GHz）太赫兹信道的时间自相关函数和空间互相关函数，太赫兹信道的相关函数如图 2 所示。可以看到，由于终端的长时间移动，不同时刻的相关函数显著不同，体现出时域非平稳性。同时，随着频率的增加，同一时刻信道的时间自相关函数和空间互相关函数将减小。

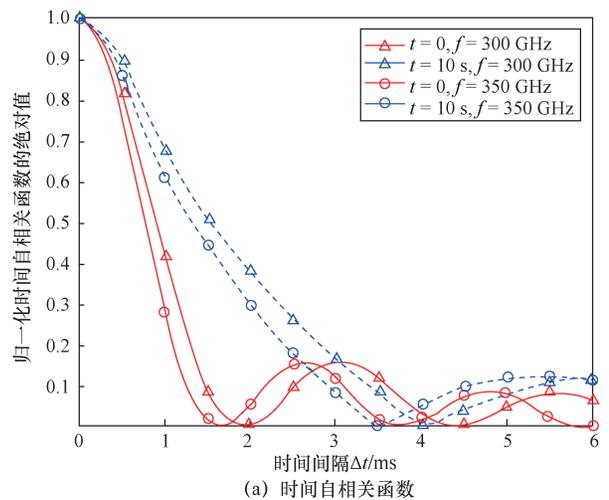

(a) 时间自相关函数

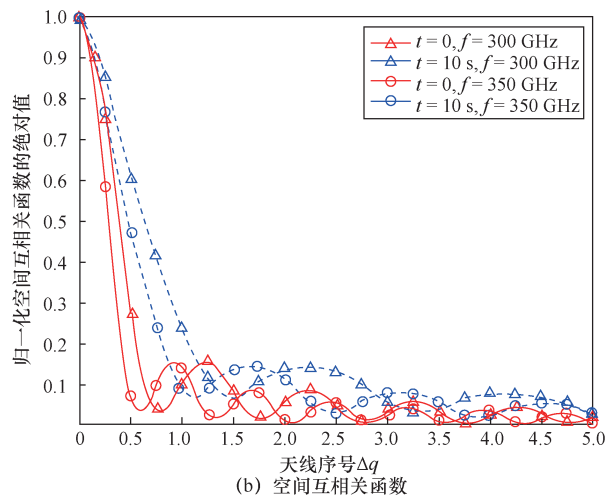

(b) 空间互相关函数

图 2　太赫兹信道的相关函数

## 3.2 光波段信道模型

对光波段信道而言，相关文献提出的确定性模型包括递归模型、迭代模型、DUSTIN 算法、天花板反射模型以及几何确定信道模型；提出的随机性模型包括几何随机信道模型与非几何随机信道模型。文献[17]对光波段信道模型进行了详细描述。

针对可见光信道，使用几何随机建模方法，提出了一种适用于车联网的可见光街角信道模型，如图 3 所示。假设在上方的 $W_1$ 建筑物有 $N_1$ 个有效散射体，用 $S^{(n_1)}(n_1=1,\cdots,N_1)$ 表示，对应的离开角和到达角分别为 $\phi_T^{n_1}$、$\theta_R^{n_1}$。在右方的 $W_2$ 建筑物有 $N_2$ 个有效散射体，用 $S^{(n_2)}(n_2=1,\cdots,N_2)$ 表示，对应的离开角和到达角分别为 $\phi_T^{n_2}$、$\theta_R^{n_2}$。将道路中间移动的行人和车辆建模为 $N_3$ 个有效散射体，用 $S^{(n_3)}(n_3=1,\cdots,N_3)$ 表示，$S^{(n_3)}$ 具有大小和方向随时间变化的速度，对应的离开角和到达角分别为 $\phi_T^{n_3}$、$\theta_R^{n_3}$。发射端移动方向与速度分别为 $\gamma_T$、$v_T$。$h_{T_1}$ 和 $h_{T_2}$ 分别表示发射端到 $W_1$ 和 $W_2$ 的距离，$h_{R_1}$ 和 $h_{R_2}$ 分别表示接收端到 $W_1$ 和 $W_2$ 的距离，$h_{31}$、$h_{32}$ 分别表示移动散射体 $S^{(n_3)}$ 到 $W_1$ 和 $W_2$ 的距离。

考虑从发射端经 $S^{(n_1)}$、$S^{(n_2)}$、$S^{(n_3)}$ 到达接收端的 3 条单次反射路径分别表示为 $SB_1$、$SB_2$、$SB_3$。可见光信道的信道冲激响应如图 4 所示，仿真参数为 $h_{T_1}=3\text{ m}$，$h_{T_2}=60\text{ m}$，$h_{R_1}=40\text{ m}$，$h_{R_2}=3\text{ m}$，$h_{31}=8\text{ m}$，$h_{32}=8\text{ m}$，$\gamma_T=0$，$\gamma_R=\pi/2$。可以看到，$SB_3$ 分量的时延较大，传播距离较远，导致接收光功率远低于 $SB_1$ 和 $SB_2$。

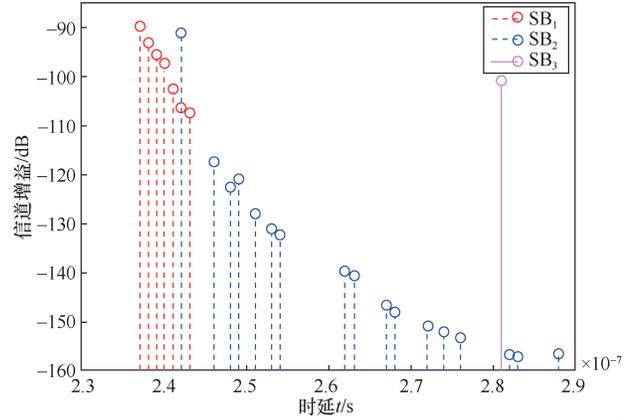

图 4 可见光信道的信道冲激响应

## 3.3 卫星通信信道模型

卫星通信信道主要为直射路径，因此，接收信号强度一般较稳定，除非受天气条件与对流层闪烁的影响。当前多数卫星通信信道模型主要考虑接收信号幅度的概率密度分布，根据接收到的信号强度，信道状态可以分为好、中等、差，从而使用马尔可夫链进行建模。同时，一些初步的研究工作尝试使用几何随机信道模型对卫星通信信道进行建模。

文献[45]提出了一种适用于高纬度地区 Q 波段的卫星通信三维信道模型，散射簇位于半球球面上，考虑接收端的运动对接收信号的影响以及受环境散射影响的莱斯因子。

仿真得到的卫星通信信道的时间自相关函数如图 5 所示。仿真频率为 $f=39.402$ GHz，接收端移动速度为 $v=3$ m/s，散射簇的移动速度为 $v_c=0.5$ m/s。理论模型与仿真模型得到的结果一致，验证了模型的正确性。在不同时刻（$t=0, 10$ s）的自相关函数不同，表明卫星通信信道具有时域非平稳性。

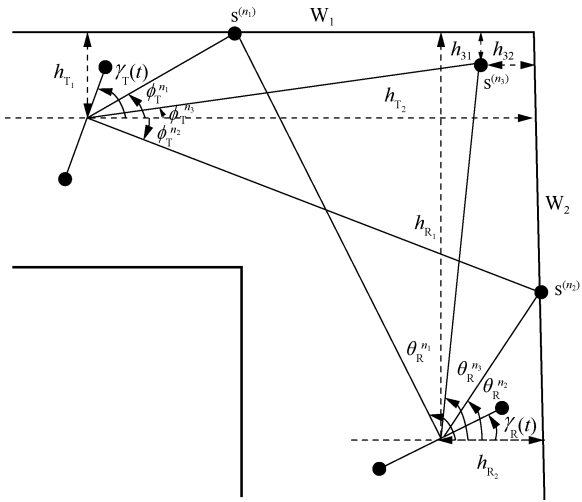

图 3 车联网可见光街角信道模型

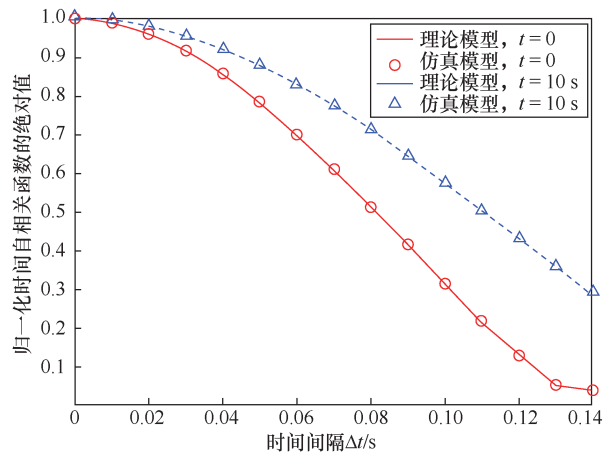

图 5 卫星通信信道的时间自相关函数

### 3.4 无人机通信信道模型

文献[20]对空—地无人机通信信道大尺度路径损耗模型进行了全面总结。无人机小尺度多径信道模型包括确定性与随机性信道模型。确定性模型包括射线追踪模型和分析性模型，如两径模型等。随机性信道模型包括规则形状几何随机信道模型、非规则形状几何随机信道模型、非几何随机信道模型与马尔可夫模型。

由于无人机通信信道具有高移动性、俯仰角大、散射体在近地空间分布等特性，使用几何随机建模方法提出了一种三维非平稳无人机空—地信道模型，详细的理论推导参见文献[46]。在规则随机几何建模的基础上引入散射簇在时间上的演进，根据几何关系实时更新信道参数，从而较好地反映无人机通信信道的时域非平稳特性。考虑散射体分布在近地空间这一特性，该模型支持根据建筑物和地形的高度限定散射体分布的高度。此外，通过改变控制簇生成算法的参数，如时延分布、角度分布等，该模型可以适用于不同的应用场景，具有较好的通用性。仿真与测量的无人机通信信道均方根时延扩展的累积分布函数如图 6 所示，结果显示，该模型能够较好地拟合开阔场景和居民区场景的均方根时延扩展。不同速度下无人机通信信道平稳间隔的概率分布如图 7 所示，结果显示，在无人机发送端速度增加（$v_T$= 5 m/s, 10 m/s, 15 m/s）、接收端速度不变（$v_R$ = 3 m/s）的情况下，平稳间隔逐渐变小，与实际情况相符。

### 3.5 海洋/水声信道模型

射线追踪模型可以用于海洋及水声信道的确定性信道仿真。除此之外，两径模型与三径模型也较为常用。随机性模型包括几何随机信道模型与两

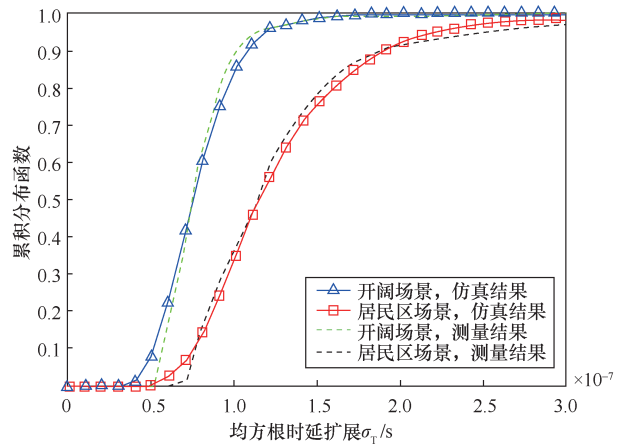

图 6 仿真与测量的无人机通信信道均方根时延扩展的累积分布函数

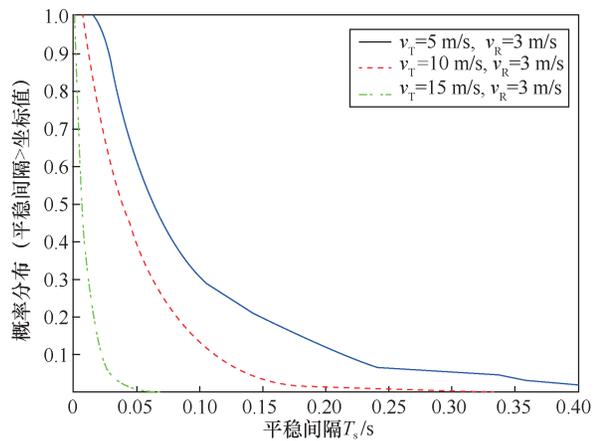

图 7 不同速度下无人机通信信道平稳间隔的概率分布

径衍射功率模型。瑞利、莱斯、对数正态分布常用于水声信道的信号幅度建模中。

针对海洋通信中典型的船—岸基信道进行建模，海洋通信船—岸基信道模型如图 8 所示，详细的理论推导参见文献[47]。在此场景中，沿岸建筑

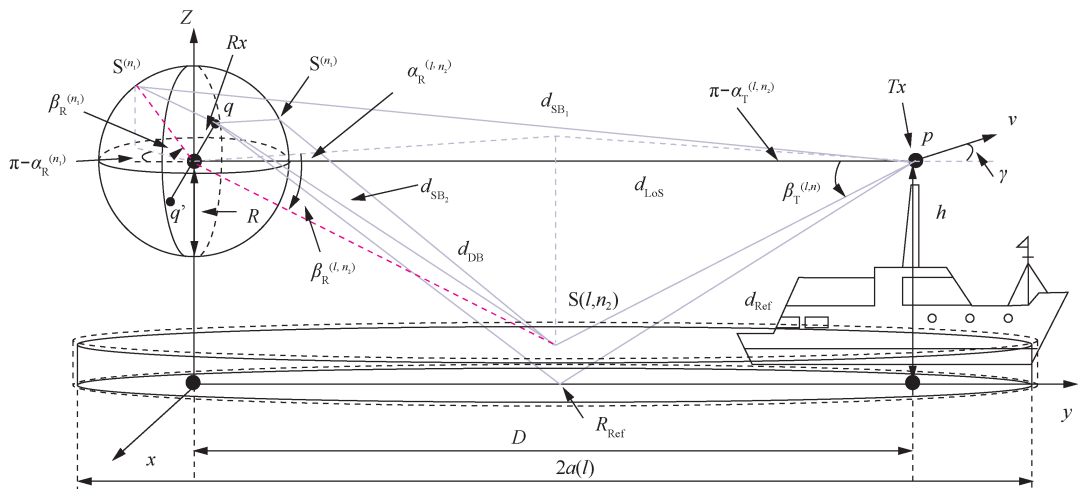

图 8 海洋通信船—岸基信道模型

物造成的散射是不可忽视的。为了描述此种散射，假设一部分散射体分布在一个以接收天线（安装在海岸附近）为中心的球面上。由于海面是起伏的表面而非静止的水平面，因此，采用多个低高度共焦椭圆柱面来描述粗糙海面的散射体分布。为了表示不同范围海洋表面波动的影响，这些椭圆柱体将采用不同长度的长轴。发送天线安装在船上，随船移动，采用单发双收天线配置。信道冲激响应是直射分量、反射分量和散射分量的组合。反射分量包括来自球面模型和椭圆柱体模型的单次反射和二次反射。仿真得到的不同莱斯因子（$K=1,2,3$）下的空间互相关函数，海洋通信船—岸基信道的空间互相关函数如图 9 所示。由图 9 可知，当 $K$ 值增加时，信道具有较强的直射径，子信道之间的空间相关性加强。

在空—时—频域的演进过程，提出一种三维高铁无线通信信道模型，详细的理论推导参见文献[48]。通过参数计算方法对角度进行离散化，提出了相应的仿真模型。基于所提高铁通信信道模型，研究高铁通信信道的统计特性。高铁通信信道的时间自相关函数如图 10 所示。由图 10 可以看到，在相同时刻，仿真模型与理论模型可以很好地匹配，且在不同时刻（$t = 2$ s, 4 s），仿真模型与理论模型的时间自相关函数差别很大，证明了所提模型可以很好地描述高铁通信信道的非平稳特性。

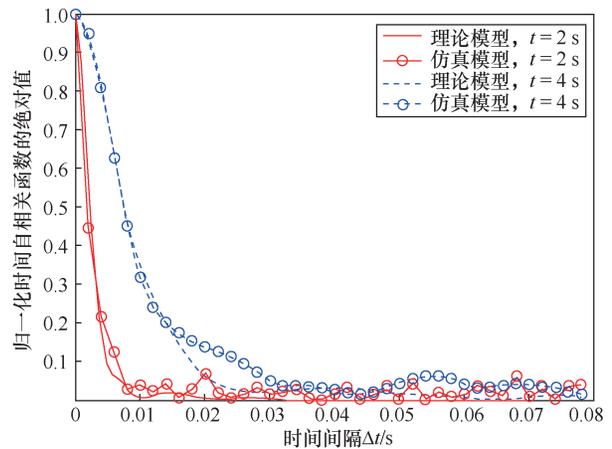

图 10　高铁通信信道的时间自相关函数

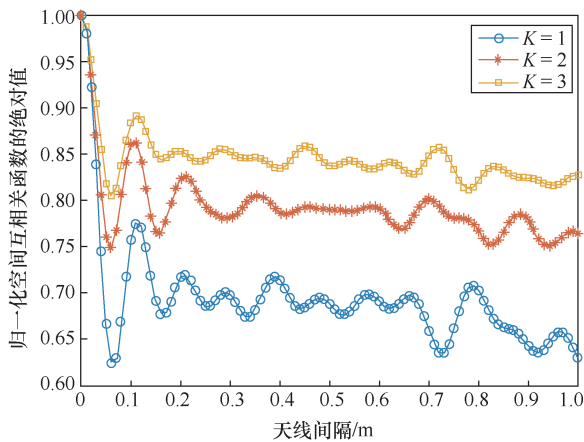

图 9　海洋通信船—岸基信道的空间互相关函数

### 3.6 高铁/车对车通信信道模型

在高铁与车对车通信信道中，需要考虑高移动性与非平稳性，文献[25]对高铁通信信道模型进行了总结。射线追踪可以用于仿真高铁与车对车通信信道中，随机性模型包括几何随机信道模型、QuaDRiGa 模型、动态模型、马尔可夫模型、传播图模型。

随着高铁用户数不断增加，通信需求量不断增长，现有的铁路通信系统无法很好地满足高带宽、低时延、高可靠性等方面的需求。未来高铁无线通信系统中需要采用大规模/超大规模天线和毫米波等新型关键技术。列车的高速运行使得高铁无线通信信道呈现较大的多普勒频移和时域非平稳特性，此外，新技术的引入将带来空域和频域的一些新的信道特性。为了准确地描述这些信道特性，考虑簇

针对车对车通信信道，建立基于孪生簇的信道模型，即信道的簇在传输环境中成对出现，表现为第一跳簇和最后一跳簇。多径的离开角和到达角分别由第一跳簇和最后一跳簇决定，且中间的传输抽象为虚拟链路。当虚拟链路的时延为 0 时，多跳径退化为单跳径。当前的车对车通信信道模型大多假设发送端/接收端以恒定速度直线运动。提出的信道模型考虑了更普遍的车对车通信场景，模型的发送端、接收端及散射体的运动速度和运动方向均可随时间变化，详细的理论推导参见文献[49]。因此，该模型能够更真实地体现车对车通信信道的时间非平稳性。

车对车通信信道的相关函数在不同时刻（$t = 0$, 1 s, 2 s）的绝对值如图 11 所示。其中，发送端和接收端的初始速度均为 30 km/h，加速度为 1 m/s，角速度为 $\pi/5 \text{ s}^{-1}$，第一跳簇和最后一跳簇以 30 km/h 的速度直线运动。可以看到，由于发送端/接收端与散射体的相对运动，模型的时间自相关函数值随时间降低。与时间自相关函数不同，该模型的空间互相关函数随时间逐渐增大，其值随时间的变化由时变的到达角和天线阵列朝向决定。

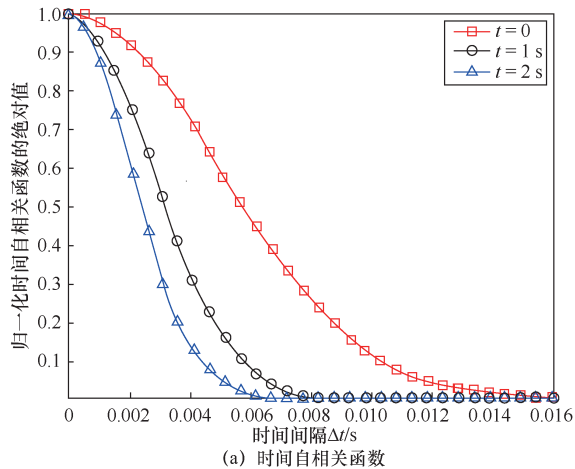

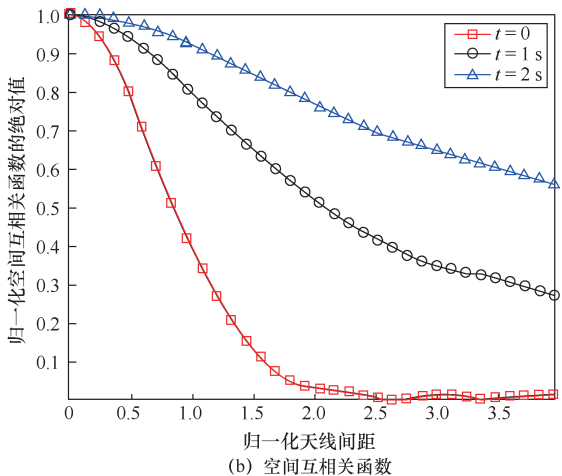

图 11 车对车通信信道的相关函数在不同时刻
（$t = 0, 1\ \text{s}, 2\ \text{s}$）的绝对值

### 3.7 大规模/超大规模天线信道模型

对于大规模/超大规模天线信道，需要考虑球面波、非平稳、簇的生灭等特性。通常，球面波可以通过在几何随机信道模型中计算每个天线单元准确的传播距离来建模。非平稳特性可以用可见域和生灭过程来建模。大规模/超大规模天线三维双簇非平稳信道模型的仿真结果[50]如图 12 所示。从图 12 可以看到，随着滑动窗在天线阵列轴上滑动，多径信号在阵列轴上呈现生灭特性，并且由于球面波前，多径的到达角在阵列轴上出现漂移。

### 3.8 轨道角动量信道模型

目前，轨道角动量信道的研究主要集中在直射场景下的信道容量推导与性能分析。文献[38]在 28 GHz 频段对采用不同传输模式的轨道角动量信道的多径效应开展了实验与仿真研究。文献[51]对基于轨道角动量的无线通信进行了概述，讨论了轨道角动量的信号产生、波束收敛、信号接收。目前的轨道角动量信道测量结果较少，主要是验证轨道角动量

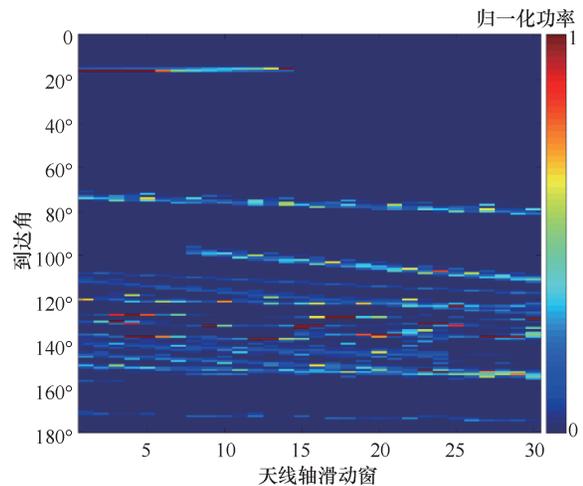

图 12 大规模/超大规模天线三维双簇非平稳信道模型的仿真结果

在不同应用场景的可行性，尚未提出基于实测结果的轨道角动量信道模型。仅有文献[52]提出了一种基于均匀圆阵的轨道角动量稀疏多径信道模型，并对不同模式的轨道角动量信道增益进行了仿真。

### 3.9 工业物联网信道模型

文献[41]总结对比了不同的工业物联网路径损耗模型，包括自由路径损耗模型、单斜率模型、3GPP 模型（RMa/Uma/UMi/InH 场景）、工厂室内模型以及全局路径损耗模型。自由路径损耗模型作为基准。单斜率模型使用传输功率与路径损耗指数描述信号强度，3GPP 模型在 RMa、Uma、UMi 和 InH 这 4 种场景下使用不同的信道模型。工厂室内模型基于大量的信道测量结果。全局路径损耗模型将直射/非直射场景考虑在内，从而更好地对信道的波动状态进行描述。

针对工业物联网信道，基于通用 5G 信道模型进行建模[53]。在该模型中，将大规模/超大规模天线简化为一般多天线，发送端、接收端、散射体均假设为缓慢移动以符合实际工业物联网场景特点。仿真了大型工厂环境下忙时与闲时的均方根时延扩展的累积分布函数，工业物联网信道均方根时延扩展的累积分布函数如图 13 所示。收发端距离设为 100 m，对视距、轻度非视距、阻挡非视距 3 种情况下的均方根时延扩展进行比较。在忙时，如工作时段，工厂内有一定数量的工人在工作，散射体数量较多，且可能处于运动状态；在闲时，如休息时段，工厂内没有工人工作，散射体数量相对较少。可以看到，忙时的散射体较多，引起多阶反射，多径具有较大的时延，从而在时延域引起信道弥散，具有较大的均方根时延扩展。由于多径反射的影

响，在视距向非视距转变的过程中，时延扩展变大，忙时与闲时的时延扩展差也进一步加大。

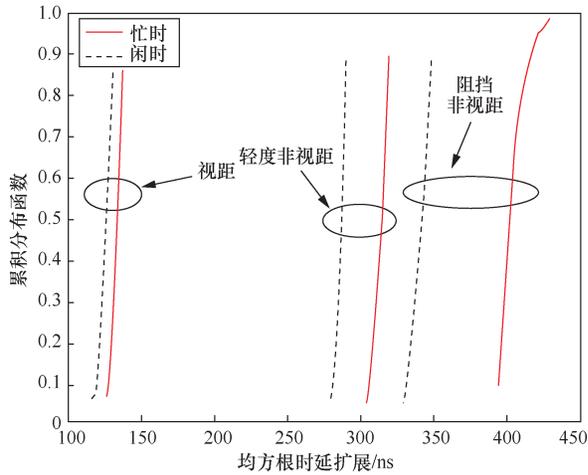

图 13　工业物联网信道均方根时延扩展的累积分布函数

# 4　面向 6G 的信道研究的未来挑战

## 4.1　6G 无线信道测量

信道测量通常需要耗费大量的人力和物力，但它仍然是对新频段和新场景下的无线信道进行研究的直接手段。通过信道测量实验，可以获取海量的信道测量数据，并通过高精度多径参数估计算法对测量数据进行处理，提取信道参数，从而对信道特性进行深入的研究与分析。但是，信道测量所使用的信道探测器因测量频段与场景不同，在测量原理、通道数、频段、带宽、发射功率、动态范围、移动性等方面具有较大差异。如何从成本、开发周期、性能参数等方面对适用于不同频段与场景的信道探测器进行综合考虑，进而高效地开展信道测量实验，是未来 6G 无线信道研究的基础。

## 4.2　基于智能反射面的 6G 技术的信道测量与模型

智能反射面是最近提出的超越大规模/超大规模天线的概念，通过集成大量低成本的无源反射元件，智能地重新配置无线传播环境，从而显著提高无线通信网络的性能[54]。智能反射面可以构成大规模/超大规模天线阵列，并可在人工智能与机器学习的帮助下，通过可重构处理网络加以控制。由于无线信道变得智能与可重构，智能反射面在满足未来 6G 通信系统需求方面展现出巨大的前景。基于智能反射面的 6G 技术的信道测量与建模对智能反射面的功能验证与性能评估必不可少。然而，目前基于智能反射面的 6G 技术的信道测量与建模工作基本空白，亟需开展相关研究。

## 4.3　人工智能使能的信道测量与模型

由于新的频段、场景的出现及天线数量的大幅增加，信道测量的数据量急剧增加，从而难以使用传统的数据处理方法。一些初步的研究工作展现了人工智能与机器学习在无线信道测量与建模方面的前景，如使用聚类、分类及回归算法分别进行信道多径分簇、信道场景分类、信道特性预测[55]。不同的机器学习算法，如人工神经网络、卷积神经网络、生成对抗网络等都可以用于无线信道建模中。与传统信道建模方法相比，应用人工智能与机器学习的一个显著优点是能够部分预测未知场景、未知频段及未来时刻的无线信道特性。

## 4.4　6G 通用标准化信道模型

在 5G 及之前的通信系统中，标准化信道模型偏向于通用的基于几何的信道模型框架，对不同的场景使用不同的参数加以描述。文献[53]提出了一种通用的三维非平稳 5G 信道模型，适用于大规模/超大规模天线、高铁、车对车、毫米波信道场景。目前，所有的标准化信道模型都只针对陆地通信网络，支持的频段仅到毫米波频段。此外，基于相关的虚拟信道表示及波束域信道模型在毫米波及大规模天线信道中也较为常用[56-58]。然而，未来 6G 信道将包含空天地海一体化网络，支持的频段将达到光波段频段。因此，推导新的 6G 通用信道模型框架将更具挑战性。由于 6G 无线信道的异构特性及无线电波的不同尺度表现，如何使用 6G 通用信道模型框架对 6G 无线信道进行准确描述成为一个难题。例如：如何融合射频无线频段（高达太赫兹频段）及光波段频段的不同光电信道特性，如何融合陆地场景与空天海场景的不同信道特性，如何建模二维及三维移动轨迹与速度变化等。

## 4.5　信道模型参数、信道统计特性与通信系统性能的复杂映射关系

信道模型参数包括大尺度参数（路径损耗、阴影衰落）和小尺度参数（多径幅度、时延、角度、多普勒频移等）。信道统计特性包括时间自相关函数、空间互相关函数、时延扩展、角度扩展、多普勒功率谱密度、平稳间隔等。通信系统性能包括信道容量、误码率、能效、频谱效率等指标。信道模型参数影响信道统计特性，进而影响通信系统性能。信道模型参数、信道统计特性与通信系统性能 3 者之间存在复杂的非线性映射关系。如何通过大量的信道测量实验和信道特性分析提出能够准确

描述信道特性的信道模型，使其能够通过调整模型参数来准确刻画信道统计特性及通信系统性能，是未来 6G 信道研究的一个重点。

### 4.6  6G 信道模型的性能评估

信道模型的性能评估可从准确度、复杂度、通用性 3 个方面衡量。信道模型的准确度一方面可以通过比较信道模型与信道测量的统计特性来判断，另一方面可以通过研究信道模型对通信系统性能的影响来判断。信道模型的复杂度可以通过模型的参数个数、运算量、仿真运行时间等衡量。信道模型的通用性主要考虑模型框架是否具有较大的灵活性，能否通过信道模型参数的调整适用于多频段与多场景。一个好的信道模型应当是准确度、复杂度、通用性 3 个方面的最佳折中。此外，所提模型与标准化信道模型的兼容性也是考虑的重点。

## 5  结束语

本文首先简单描述了面向 6G 的无线通信网络的愿景，包括新的应用需求、技术需求、性能指标与使能技术。然后全面论述了面向 6G 的全频谱全场景无线信道测量与信道特性的最新进展，包括毫米波、太赫兹、光波段、卫星、无人机、水声、高铁、车对车、大规模/超大规模天线、轨道角动量以及工业物联网通信信道，并展示了面向 6G 的无线信道的相关测量与建模结果。通过分析面向 6G 的无线信道测量与建模的已有研究成果，归纳研究方法，提供新的思路，从而为研究人员提供重要参考。总体而言，针对本文论述的面向 6G 的无线信道，已开展了初步的研究工作，但信道测量结果仍然较少，信道特性研究不够深入，信道模型研究仍不充分，需要在未来开展更为深入细致的研究工作。本文同时指出了未来 6G 信道测量与建模研究的挑战。

[作者简介]

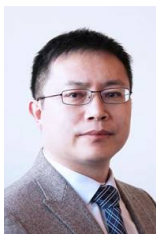
王承祥（1975- ），男，山东高密人，东南大学移动通信国家重点实验室教授、紫金山实验室兼职教授，主要研究方向为无线信道测量与建模、6G 通信网络架构和关键技术、人工智能与无线通信网络的融合等。

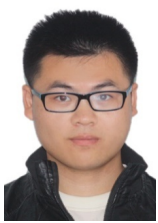
黄杰（1991- ），男，湖北黄冈人，博士，东南大学移动通信国家重点实验室博士后、紫金山实验室兼职科研人员，主要研究方向为毫米波、太赫兹、大规模/超大规模天线信道测量与建模、B5G/6G 关键技术等。

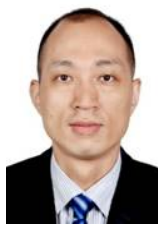
王海明（1975- ），男，江苏江阴人，博士，东南大学毫米波国家重点实验室教授、紫金山实验室兼职教授，主要研究方向为毫米波无线通信、毫米波雷达成像、无线传播测量与信道建模、多频段宽带天线与阵列等。

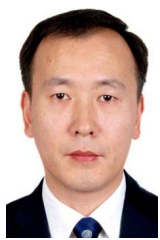
高西奇（1967- ），男，安徽灵璧人，东南大学移动通信国家重点实验室教授、紫金山实验室兼职教授，主要研究方向为宽带无线通信、大规模无线通信、无线通信信号处理等。

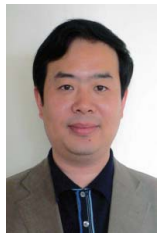
尤肖虎（1962- ），男，山东济宁人，博士，东南大学移动通信国家重点实验室主任、紫金山实验室常务副主任，主要研究方向为移动通信、自适应信号处理、人工神经网络在通信及生物医学工程中的应用等。

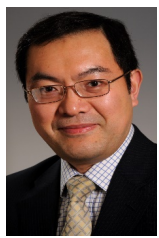
郝阳（1967- ），男，江苏南京人，博士，伦敦玛丽女王大学电子工程与计算机科学学院教授，主要研究方向为计算电磁学、微波超材料、石墨烯和纳米微波、体域网天线和无线传播、毫米波/亚毫米波有源天线、光子积体天线等。